\documentclass[aps,prb,preprint,groupedaddress]{revtex4-2}

\usepackage{graphicx}
\usepackage{dcolumn}
\usepackage{bm}

\newcommand{\vect}[1]{\mathbf{#1}}

\newcommand{\eps}{ { \varepsilon } }
\renewcommand{\phi}{{\varphi}}

\newcommand{\eff} {\mathrm{eff}}

\begin{document}

\title{A dynamical approach to the $\alpha$--$\beta$ displacive transition of quartz}


\author{Andrea Carati}
\email{andrea.carati@unimi.it}
\affiliation{Department of Mathematics, Universit\`a
    degli Studi di  Milano Via Saldini 50, 20133 Milano, Italy}
\author{Fabrizio Gangemi}
\affiliation{DMMT, Universit\`a di Brescia, Viale Europa
    11, 25123 Brescia - Italy}
\author{Roberto Gangemi}
\affiliation{DMMT, Universit\`a di Brescia, Viale Europa
    11, 25123 Brescia - Italy}
\author{Luigi Galgani}
\affiliation{Department of Mathematics, Universit\`a
    degli Studi di  Milano Via Saldini 50, 20133 Milano, Italy}

\date{\today}

\begin{abstract}
 General features of the $\alpha-\beta$ transition of quartz are
 investigated. Molecular dynamics methods are mainly used, an analytic
 treatment being deferred to a work in preparation. A basic
 preliminary observation is that the transition involves only a
 subsystem of four normal modes on which the remaining ones just act
 as a reservoir. The dynamics of the relevant subsystem turns out to
 be Hamiltonian, being governed by an effective potential that depends
 on the specific energy of the total system. The effective potential
 is actually calculated through time averages. It describes the
 transition as a pitchfork bifurcation, and also explains the
 phenomenon of the soft mode, since it exhibits a frequency that
 vanishes at the transition. The critical exponent too is estimated.
\end{abstract}


\maketitle

\section{Introduction}
A displacive phase transition occurs when a crystal passes, by
increasing temperature, from a less symmetric structure to a more
symmetric one. This takes place through small changes of the atoms'
positions, at variance with the reconstructive phase transitions in
which broader changes are met. A distinctive feature of the displacive
transitions is the phenomenon of the soft mode, i.e. the vanishing of
a characteristic frequency of the crystal at the transition.
A very large body of literature exists on the subject, as shown in the review
articles by Dove~\cite{dove1997} on displacive phase transitions in general,
and by Scott~\cite{scott} on soft modes.

Here we consider the paradigmatic case of the $\alpha-\beta$
transition of quartz, which is extensively studied in the literature,
and investigate it through molecular dynamics methods in terms of the
crystallographic normal modes. A related theoretical work is in
preparation~\cite{nostro}. Now, at the experimental level it is known
that the soft mode is present in the Raman spectra and not in the
infrared ones. Thus we study the dynamics of the four normal modes
that are active in the Raman spectrum and not in the infrared one. On
the other hand it turns out that such four modes completely describe
the transition, and for this reason they will be called here the
\textbf{relevant modes}.

It turns  out that the four relevant modes form a strongly coupled
subsystem, on which the remaining modes just act as a thermal noise
depending on the total energy. This fact, illustrated in
Section~\ref{sec:4}, is a first characteristic feature of the transition.
Apparently it was not explicitly pointed out in the literature. In fact
we happened  to observe it  occasionally, and  this even gave origin to the
present  research.

The key point, however, is that the presence of a relevant subsystem
presenting a closed dynamics allows one to identify the mechanism that
produces both the transition and the soft mode, and this is just the
existence of an energy-dependent effective potential that is found to
govern the motion of the relevant modes. Such potential is simply
determined by averaging, over all the remaining modes, the
instantaneous accelerations of the four relevant ones.  In this
connection, the averaging procedure actually used is itself
noteworthy, since it is completely different from the standard
canonical one of statistical mechanics. In fact the average is
essentially just a time average computed up to the first time at which
it stabilizes, a procedure that does not rely at all on a presumed
chaoticity property of the considered motions (in this connection, see
the remarks in \cite{anderson}).

Now the effective potential is a function of the coordinates of the
relevant modes and, when plotted along a certain direction to be
described below, it exhibits a pitchfork bifurcation, which determines
the form of the transition.
There is a certain analogy with the Landau theory of phase
transitions.  But the latter remains at a phenomenological level,
being based on a macroscopic thermodynamic function (free energy),
whereas here we are dealing, at a microscopic level, with a purely
dynamical quantity, i.e., the energy-dependent effective potential.
The effective potential also produces the soft mode, since it allows
one to identify a frequency dependent on specific energy that vanishes
at the transition.

The paper is organized as follows. In Section~\ref{sec:2} we briefly
recall the structure of the $\alpha$ and of the $\beta$ quartz in
terms of normal modes, while in Section~\ref{sec:3} are given some
details on the quartz model used in the numerical computations.  In
Section~\ref{sec:4} we show how the system breaks down into the two
subsystems, that of the relevant modes, and the reservoir, while in
Section~\ref{sec:5} we show how the transition is described in terms
of the relevant subsystem.  Finally, in Section~\ref{sec:6} the
effective potential is constructed, showing how it exhibits a pitchfork
bifurcation and the existence of a soft mode. The conclusions follow.

\section{The crystal configuration in terms of normal  modes}\label{sec:2}

The quartz crystal is a Bravais lattice with lattice vectors
$\vect a$, $\vect b$, and $\vect c$, where $\vect a$ and $\vect b$
form an angle of $2\pi/3$ with one another, whereas $\vect c$,
orthogonal to both of them, is set parallel to the optical axis.
The primitive cell contains 3 Silicon atoms and 6 Oxygen atoms. In the
$\alpha$ phase its point symmetry group is $D_3$, while in the $\beta$ phase
the point symmetry group is $D_6$, of which $D_3$ is a subgroup.
Transformations of these groups are combined to
translations along the optical axis to give the complete set of symmetries
of the structure, which may therefore be right handed or left handed.
The cell configuration is defined once the coordinates of one Oxygen
and of one Silicon atom are assigned: indeed, by applying all  symmetry
transformations to these coordinates, the positions of all the other atoms
are obtained. In particular, when the atomic positions are expressed as
$\vect x=x\vect a+y\vect b+z\vect c$, where $x,y,z$ are called the fractional
coordinates, the symmetry of the $\alpha$ phase requires that the
coordinates $y_S$ and $z_S$ of the first Silicon atom vanish, while no
constraint is put on $x_S$ and on the coordinates $x_O,y_O,z_O$ of the first
Oxygen atom. The symmetry of the $\beta$ phase, in addition to the previous
constraints, requires $x_S=1/2$, $x_O=2y_O$ and $z_O=1/6$.

It is well known that, for a given chirality, two energetically equivalent
$\alpha$ configurations exist, related by a rotation of $\pi$ about the
optical axis. In the microscopic model adopted in this work the inter atomic
potential presents indeed two configurational minima with the symmetry
properties of the $\alpha$ phase.
We consider normal modes calculated with respect to one of such minima
and describe the crystal configuration in terms of the corresponding
coordinates. When the crystal is in the chosen minimum, all mode coordinates
vanish by definition. The other possible crystal structures exhibited by quartz
are the second energy--minimum configuration, presenting $\alpha$ symmetry,
and the $\beta$ configuration, presenting a higher symmetry group, that
contains the one of the
$\alpha$ phase as a subgroup. In these configurations, only coordinates of the
totally symmetric modes ({\it i.e.} modes that preserve the original symmetry)
may differ from zero: there are four such modes among the 27 ones of
the primitive cell. Using the inter atomic potential illustrated below,
the four totally symmetric modes have calculated frequencies of 240,
425, 586, and 1097 cm$^{-1}$ (to be compared with experimental values of 207,
356, 464, and 1085 cm$^{-1}$ respectively). The corresponding
coordinates will be denoted as $A_1,A_2,A_3,A_4$.

Thus the $\alpha-\beta$ transition may only involve such modes,
and for this reason we concentrate on them, with the aim of understanding
their dynamics.

\section{The  model}\label{sec:3}

Some details are now given concerning the molecular dynamics
simulations performed.  Quartz and other crystals exhibiting
displacive phase transitions have been the subject of many molecular
dynamics studies, beginning with the early works by Tsuneyuki et
al.~\cite{giapponesi,giapponesi2}, and several force fields have been
developed~\cite{schaible} and applied to the study of such phenomena
(see {\it e.g.} \cite{cowen} for a comparison of different force
fields in the study of phase transitions).

Coming to our work, since it is
intended as having a prevalently qualitative character, the
computations were performed in the spirit of ergodic theory, i.e.,
without making any attempt at simulating realistic situations
involving fixed pressure. So we worked at fixed volume, setting the
lengths of the basis vectors of the cell at their experimental values
\cite{kihara} at standard conditions: $a=b=4.9137\ \mathring{\rm A}$,
and $c=5.4047\ \mathring{\rm A}$.  Analogously, in place of
temperature we chose as a parameter specific energy $\eps$, {\it
  i.e.}, the total energy per degree of freedom (expressed in Kelvin,
after dividing it by the Boltzmann constant).

For what concerns the microscopic inter atomic potential, whose choice
is indeed a delicate point, we just employed the well known BKS one
reported in \cite{bks}. This takes into account both the short--range
contributions of the van der Waals forces due to the electrons in the
atoms, and the Coulomb forces due to the partially ionic character of
the atoms, cared as usual through Ewald methods.
Notice that, since the atoms are dealt with as point like, we cannot
reproduce the Raman effect, which requires, as is well known, a
deformable atomic model.  However, it will be shown later how the
chosen model allows one to exhibit a characteristic feature of the
soft mode, i.e., the existence of a frequency that depends on
temperature, vanishing at the transition.

So we use a model consisting of $N$ massive points  located in a
``working domain'' of volume $V$, consisting of $4\times 4 \times 4$
primitive cells.  Since, as already recalled, each cell is constituted
by three SiO$_2$ groups, the whole working domain contains $N=9\times
4^3=576$ atoms.  The corresponding number of modes is thus 1728.  The
integration method was the standard one of Verlet (or leap-frog) with
a time step of typically 2 femtoseconds.

The initial data were assigned by setting the system in the chosen minimum
of the inter atomic potential and generating random velocities of the particles,
according to a Maxwell-Boltzmann distribution. In terms of normal modes, this
corresponds to vanishing initial coordinates.

For each value of the energy, 20 different orbits (each with a
time--length of 400 picoseconds) were  computed: half with initial
positions in the minimum chosen for defining the normal modes, and
half with initial positions in the other minimum. The averages were
always meant as time averages along each orbit (in some cases
restricted to the last 200 ps), followed by a further average over the
different orbits.
\begin{figure}[!t] 
  \begin{center}
    \includegraphics[width = 0.5\textwidth]{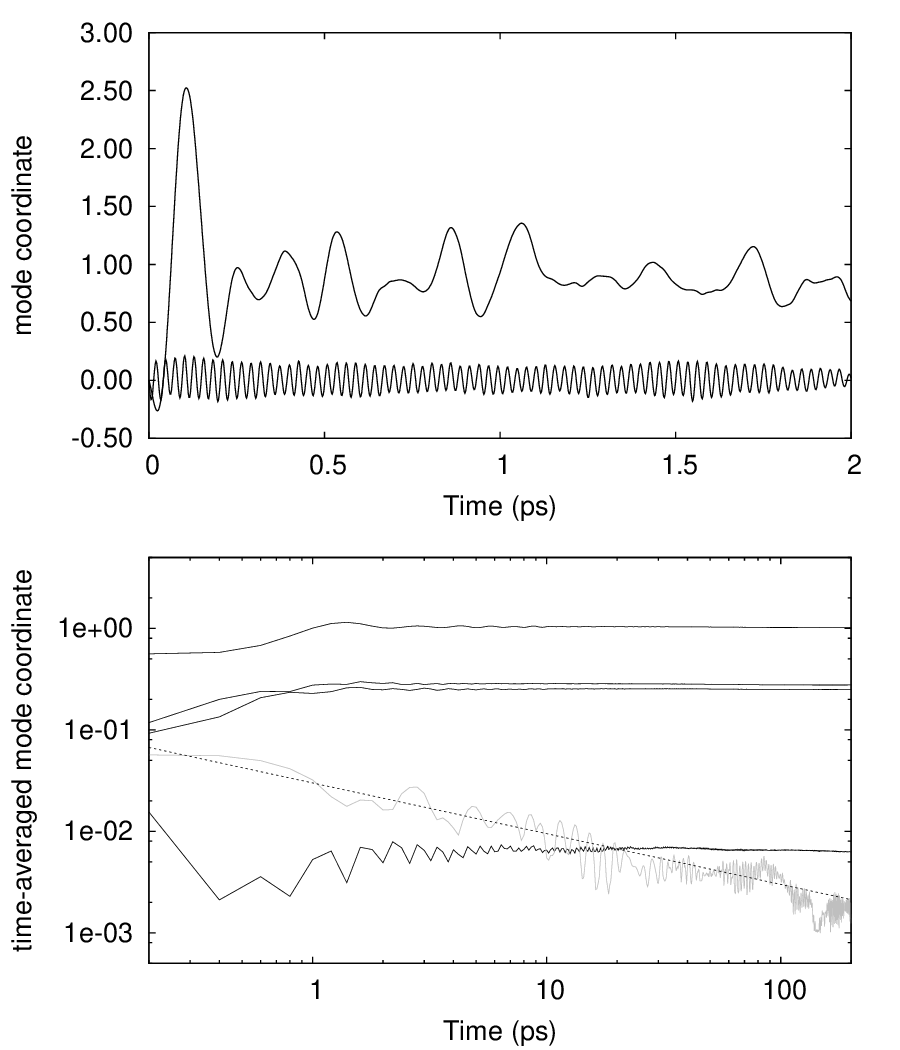}
  \end{center}   
  \caption{\label{fig:1} \textsl{Upper panel: Relevant versus noisy
      modes}. Starting from a common vanishing value of their
    coordinates, some modes jump to a finite mean value, then oscillating
    about it, and are
    relevant for the configuration of the crystal, while other ones
    keep oscillating about the initial vanishing  value, producing noise.
    \textsl{Lower panel}, time-averages in log-log scale.
    \textsl{Four (over the total of 1728) is the number of relevant
      modes (black lines)}.  Gray line is the maximum, over all the
    remaining modes, of the absolute values of time-averaged
    coordinates.  Compare with the dashed line with slope $-\,
    1/2$. Specific energy, 300 K.}
\end{figure}

\section{The decomposition into the relevant  subsystem and a
  reservoir}\label{sec:4}

The dynamical decomposition of our model of quartz into two
subsystems, mentioned in the Introduction, is vividly exhibited by
numerical simulations.  Having chosen initial data as explained above,
i.e., with vanishing values of the coordinates of all modes, and with
a certain kinetic energy (corresponding to a specific energy of 300 K,
in the present case), the subsequent evolution of the system is
illustrated in Figure~\ref{fig:1}.  In the upper panel are reported,
versus time, the instantaneous coordinates of two chosen modes, and it
is seen that, after a short transient, one of them keeps oscillating
about the initial vanishing value. Instead the other one performs an
initial jump and then oscillates (rather irregularly) about a
non vanishing value.  In the lower panel are instead reported
time-averaged coordinates (in absolute value) rather than
instantaneous values (still versus time, but in log-log scale), and
the figure allows one to see how the whole set of modes
behaves. Indeed black solid curves are reported for each of the four
relevant modes, and they are seen to stabilize at finite non vanishing
values, after a transient of the order of ten picoseconds.
For all the remaining modes, instead, a
single collective curve is sufficient to show that all their time-averaged
coordinates keep diminishing, apparently towards a vanishing value.
This is exhibited by the gray curve, which reports the maximum absolute
value of the time averages of all them.  The maximum is seen to
decrease as the inverse square root of the time over which the average
is performed (dotted line), as should be expected if, in the
$t\to\infty$ limit, the time averages actually vanished.  Thus the
motions of the coordinates of the latter modes just consist of
fluctuations about the zero value. So dynamically the system is
seen to actually be decomposed into two subsystems: the first one,
composed of the four \emph{relevant modes}, and the other one constituted
by all the remaining modes, that might be called \emph{the reservoir},
since it just acts on the relevant one as a dynamical noise.  The same
phenomenon occurs for all the specific energies investigated in our
simulations, from 300 up to 2000 K.  The phenomenon just described was
indeed for us a kind of ``little discovery'', that in fact happened to
give origin to the present research.

\section{The transition in terms of the relevant modes}\label{sec:5}

As specific energy is increased above 300 K, it is found that the
time-averaged coordinates of the four relevant modes increase (in
modulus), with the system still remaining in the $\alpha$
structure. Instead, if energy is sufficiently raised, the system
exhibits a transition to the $\beta$ phase.

The transition is neatly exhibited by means of a suitable order
parameter that we defined (through relation~(\ref{eq:eta}) - see
below) in such a way that it vanishes in the $\beta$ phase
(analogously to what occurs for the ferromagnetic transition in terms
of magnetization), while being equal to 1 in the extreme $\alpha$
phase, i.e., in the two minima.  And in fact in the upper panel of
Figure 2, which reports the order parameter $\eta$ versus specific
energy, the transition is seen to occur at a specific energy of about
1500 K (recall that we are working at fixed volume rather than at fixed
pressure).
\begin{figure}[t]
 \begin{center}
  \includegraphics[width = 0.5\textwidth]{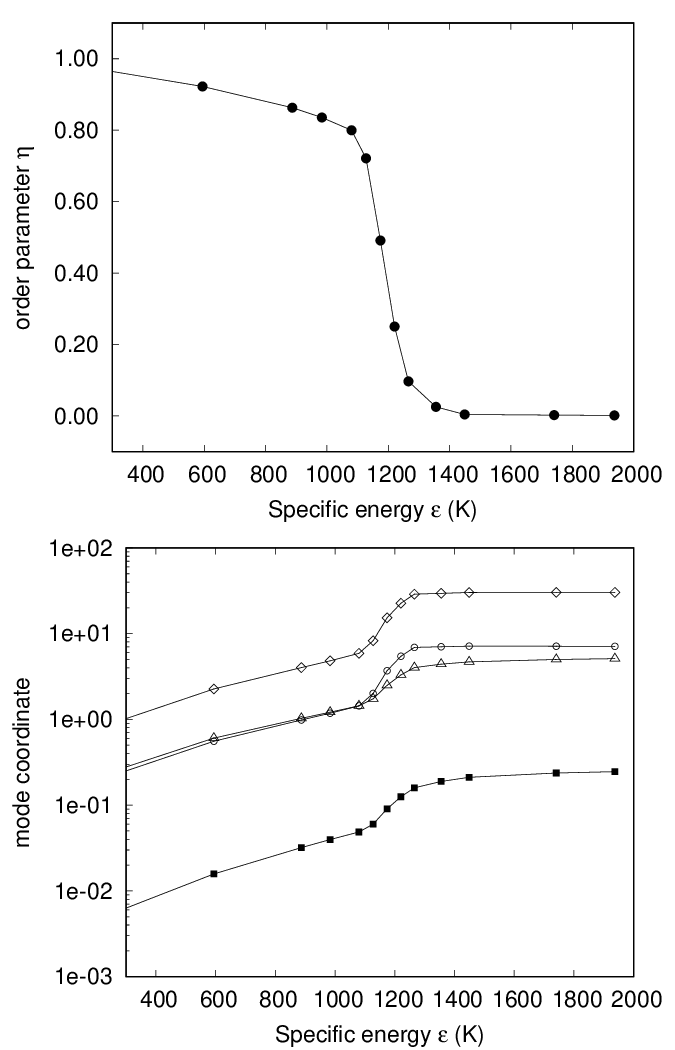}
 \end{center} 
 \caption{\label{fig:2} Time averages versus specific energy $\eps$.
     Upper panel: order parameter $\eta$ (see text). Lower
   panel:  coordinates  of the four relevant modes
   (in semi logarithmic scale).}
\end{figure}

The order parameter was defined in the following way.  Inspired by the
parameters considered in the literature, we chose it in a form
measuring the ``distance'' of a generic configuration from the $\beta$ one.
Such distance, $\eta$, can be expressed in terms of
the fractional coordinates that define, as mentioned in Section \ref{sec:2},
the structure of the cell, namely the $x_S$ coordinate of the first
Silicon and the coordinates $(x_O,y_O,z_O)$ of the first Oxygen, as the
following ratio:
\begin{equation}
\label{eq:eta}
  \eta = \frac {\sqrt{(2x_{S}-1)^2+(2y_{O}/x_{O}-1)^2+(6z_{O}-1)^2}}
       {\sqrt{(2x^\alpha_{S}-1)^2+(2y^\alpha_{O}/x^\alpha_{O}-1)^2+(6z^\alpha_{O}-1)^2}}\ ,
\end{equation}
where $x^\alpha_S$ and $(x^\alpha_O,y^\alpha_O,z^\alpha_O)$ are the
coordinates in the equilibrium configuration.  Indeed it is known that
in the $\alpha$ phase the four above mentioned coordinates are the
only free parameters, whereas in the $\beta$ phase they have to
satisfy the three conditions $x_S=0.5$, $x_O=2y_O$, and $z_O=1/6$,
i.e., the condition $\eta=0$. Thus, for $\eta\ne 0$ the crystal is in
the $\alpha$ phase, and in particular the value $\eta=1$ is attained
at the two equilibrium configurations.

But the transition may as well be displayed directly in terms
of the coordinates of the relevant modes. In fact  in the lower panel of the
same Figure 2 are shown, still versus specific energy, the time
averages of the coordinates (in absolute value) of the four
relevant modes. This figure exhibits, for all such modes, a behavior
correlated to that of the order parameter. Indeed, after a moderate
increase at lower energies, such averaged coordinates present an
abrupt change of slope at about  1100 K, and then become almost constant
starting from  about 1500 K. This should correspond to the value of
specific energy at which the $\alpha-\beta$ phase transition
macroscopically occurs.  Particularly relevant is the fact that, at the
transition, the coordinates of the two dominant  modes appear to have
attained values which remain constant for larger specific energies, which
is a signature of a non-analytic behavior characterizing the
occurring of a phase transition. Moreover, this turns out to occur already for a
finite number $N$ of atoms, without any need of attaining the
thermodynamic limit.
\begin{figure}
  \begin{center}
    \includegraphics[width=0.5\textwidth]{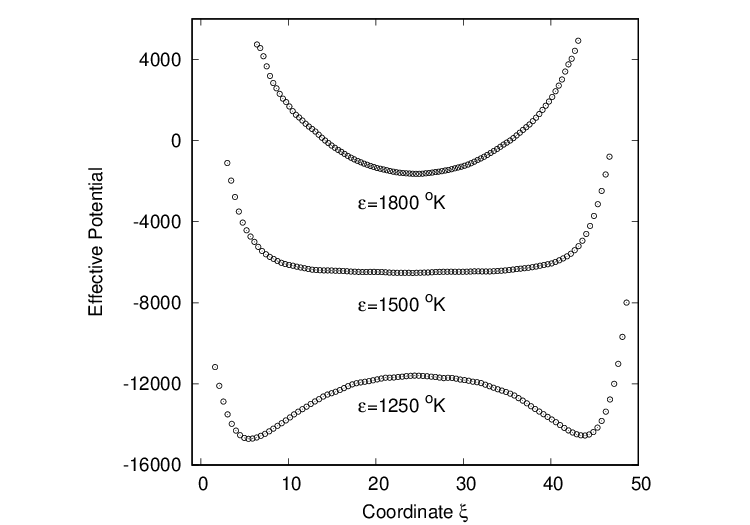}
  \end{center} 
  \caption{\label{fig:3} Plot of the numerically computed effective
    potential versus the coordinate $\xi$ along the segment joining
    the two minima of the inter atomic potential, for three values of
    specific energy, $\eps=1250$ K, $1500$ K and $1800$ K.  Notice
    that additive terms depending on specific energy were added.}
\end{figure}

\section{The energy-dependent effective potential and the soft mode
  phenomenon}\label{sec:6}

The results illustrated in Section~\ref{sec:4} about the decomposition
of the global system into the relevant subsystem and a reservoir,
suggest that a reduced dynamics may exist for the subsystem of
relevant modes, being governed by an energy-dependent effective
potential $V_\eff(\vect A,\eps)$, where we have denoted by $\vect A= (A_1,$$
A_2, A_3, A_4)$ the coordinates of the relevant modes.  The existence
of such potential is proven in a theoretical work of ours (still in
preparation), by suitably averaging the accelerations produced on the
relevant modes by all the other ones of the global system, making use
of suitably adapted methods of Statistical Mechanics. An analytic
expression of the potential is also available. Here we just limit
ourselves to assume that the potential exists, and compute
it numerically through time averages.\footnote{However, the  fact that a
  concrete computation is actually possible, on the basis of time averages
  performed over different  lapses of time, is by itself an indication that the
  result is consistent with the existence of a potential. } 
\begin{figure}[!t]
  \begin{center}
    \includegraphics[width=0.5\textwidth]{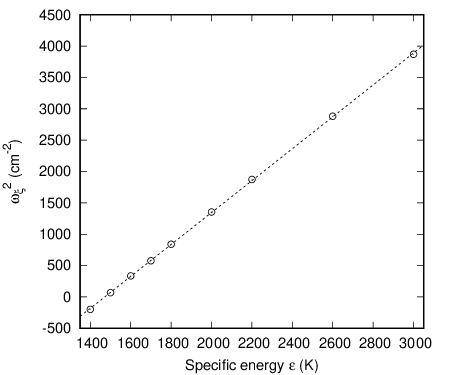}
  \end{center}
  \caption{ \label{fig:4} Squared characteristic frequency (in cm$^{-2}$)
   associated to the variable $\xi$, as a function of specific energy,
   calculated as described in the text. The dotted line represents an
   interpolation obtained by linear regression $\omega_\xi^2(\eps)=a\eps+b$.}
\end{figure}

The effective potential is computed along the direction that, in the
space $\vect A= (A_1, A_2, A_3, A_4)$ of the relevant modes, joins the
two points of minimum of the microscopic potential.  Denoting by $\xi$
the coordinate along such axis, i.e., the scalar product $\xi=\vect A
\cdot \vect e$ where $\vect e=(0.9669,-0.2375,-0.0927,0.0019)$ is the
unit vector defining such direction, the computation is performed in
terms of time-averages in the following way.\footnote{Such method is
analogous to the one we already used for computing the effective
potential acting among the two protons in the ion of the Hydrogen
molecule - see \cite{H2+}. Interestingly enough, in that case the
potential, still defined through time averages, turned out to exist
only in presence of ordered, rather than chaotic, motions. On the
other hand, ordered motions are met also in the present case, just in
virtue of the definition itself of a displacive transition.}  Given a
value of $\xi$ attained along a given orbit at a certain time, one can
compute by elementary formulas the corresponding acceleration
$\ddot{\xi}$. Thus, by collecting all such pairs from all simulations
at a given specific energy, a large sample of $(\xi,\ddot{\xi})$ pairs
is obtained.  The acceleration $\ddot{\xi}$ as a function of $\xi$ is
then determined by dividing the range spanned by the collected values
of $\xi$ into a certain number (100) of bins and calculating the
average values of $\ddot{\xi}$ over all points with $\xi$ in each
bin.\footnote{In other terms, one is thus computing the acceleration
as the conditioned mean for a given value of $\xi$, in analogy with
the procedure followed in the theoretical paper in
preparation. However, while in the theoretical paper phase-space
averages are performed, in the present case, due to the way in which
the samples are collected - i.e., through numerical solutions of the
equations of motion -, it should be clear that we are estimating time
averages.}  By further integrating with respect to $\xi$, the
effective potential is then easily calculated (apart from an additive
constant).

The results are shown in Figure~\ref{fig:3}, for three values of 
specific energy, i.e., $\eps=1250$ K, $1500$ K and $1800$ K. The figure
clearly shows that the effective potential exhibits a pitchfork
bifurcation when the system passes from the high-energy phase to
the low-energy one. Moreover, the potential is seen to be
extremely flat at the transition, a fact that should correspond to the
occurrence of a soft mode.

The shape of the effective potential at high energy naturally suggests
to associate a characteristic frequency to the variable $\xi$, namely,
the frequency of the small oscillations about the minimum, here
denoted as $\omega_\xi(\eps)$. In order to calculate this quantity, at
each value of $\eps$ the curve of the acceleration $\ddot\xi$ as a
function of $\xi$ was analyzed in the region $15<\xi<35$, where it
turns out to be linear, and its slope, computed by linear regression,
was identified with $-\omega_\xi^2(\eps)$. The results are reported in
Figure~\ref{fig:4} for specific energies in the interval $[1400,3000]$
K. Such figure is the analogue of Figure~10 of the review
\cite{scott}, taken from \cite{shirane}, obtained by inelastic neutron
scattering.  As is clearly shown by the superimposed linear
interpolation (dotted line), $\omega_\xi^2(\eps)$ has a linear
dependence on $\eps$, and vanishes at $\eps=\eps_{cr}\simeq 1471$ K.
The latter value can be interpreted as an estimate of the critical
specific energy in our model, i.e., the specific energy at which the
$\alpha-\beta$ transition occurs, since the effective potential passes
from the double-minimum to the single-minimum form. The characteristic
frequency associated to $\xi$ is thus seen to pass through the zero
value when the transition energy is attained from above. Notice also
that the linear dependence of $\omega_\xi^2(\eps)$ on specific energy
implies a relationship $\omega_\xi(\eps)\sim\sqrt{\eps-\eps_{cr}}$, so
that the critical exponent has value 1/2. As in this range of energies
the temperature should be proportional to specific energy $\eps$
(apart from an additive constant), this result is in agreement with
the experimental data. Concerning the spectrum below the critical
energy, we don't yet have sufficient elements, and just limit
ourselves to a short comment in the conclusions.

\section{Conclusions}\label{sec:8}
Summarizing, the following two phenomena were observed.

\begin{enumerate}
\item The transition has a dynamical origin. The key dynamical element
  is the effective potential which governs the four relevant modes
  that turn out to fully describe the transition. Such potential is
  energy-dependent and presents a pitchfork bifurcation, at a critical
  value of about 1500 K (working at fixed volume rather than
  pressure).  At lower energies the effective potential presents two
  symmetric minimum points, at a value that depends on specific
  energy, and the system is in an $\alpha$ phase.  Above the critical
  energy the potential presents just one minimum, that is independent
  of specific energy, and the system is in a $\beta$ phase.  A
  theoretical explanation of such a general description will be given
  in our work in preparation.
 
\item For energies higher than the critical energy there exists a
  frequency that vanishes at the transition with an exponent equal to
  1/2, in agreement with experiments.  Now, the analytical reason for
  such a behavior was not yet understood, and is still under
  investigation. We are however confident that such feature should be
  the origin of the soft mode observed in the experimental data.
  Concerning the behavior below the transition, where the critical
  exponent is experimentally near to 1/3~\cite{scott}, it should be
  recalled at least that, as is well known, the value of such
  frequency depends on density.  So we cannot fully discuss such case
  in our model, which is formulated for fixed density.
\end{enumerate}

\indent
In conclusion, the main point made in the present paper is that the
description of displacive transitions should be performed in terms of
the (energy-dependent) effective potential governing the relevant
modes, rather than in terms of the original microscopic inter atomic
potential, from which the effective potential can be deduced.

An aside remark is that the present result may be of interest for the
more general problem of a statistical mechanics explanation of phase
transitions.  The point is that the transitions is determined by only
four modes, irrespective of the number of atoms constituting the
system, either finite or infinite, whereas it is often maintained that
an infinite number of constituents would be necessary.

\begin{acknowledgments}
 F. Gangemi wishes to thank the HPC department of CINECA for access
 to computing resources.  A.~Carati, L.~Galgani and F.~Gangemi
 performed this work in the framework of GNFM activities.
\end{acknowledgments}

\bibliography{soft-PRB}

\end{document}